\documentclass[12pt,nofootinbib]{article}

\usepackage[dvips]{graphicx}
\usepackage{amssymb}
\usepackage{amsmath}
\usepackage{epsfig}
\usepackage{amsfonts}
\usepackage{color}
\usepackage{mathtools}
\usepackage{enumerate}

\definecolor{White}{rgb}{1,1,1}
\definecolor{Black}{rgb}{0,0,0}
\usepackage{sidecap}
\usepackage{epsf}
\usepackage{graphicx,epsfig}
\usepackage{amsfonts}
\usepackage{amssymb}
\usepackage{tabularx,latexsym,alltt,graphicx,textcomp,hyperref,amsmath,amssymb,color}
\makeatletter
\renewcommand\section{\@startsection {section}{1}{\z@}%
                                 {-3.5ex \@plus -1ex \@minus -.2ex}
                                   {2.3ex \@plus.2ex}%
                                   {\normalfont\large\bfseries}}
\renewcommand\subsection{\@startsection{subsection}{2}{\z@}%
                                   {-3.25ex\@plus -1ex \@minus -.2ex}%
                                     {1.5ex \@plus .2ex}%
                                     {\normalfont\bfseries}}
\renewcommand\subsubsection{\@startsection{subsubsection}{3}{\z@}%
                                   {-3.25ex\@plus -1ex \@minus -.2ex}%
                                     {1.5ex \@plus .2ex}%
                                     {\normalfont\itshape}}
\makeatother

\newcommand{\Letter}{
\setlength{\textwidth}{16.5cm}
   \setlength{\textheight}{22.8cm}
    \hoffset=-0.5in
\voffset=-2.1cm }

\Letter

\newcommand{\bra}[1]{\left\langle #1 \right|}
\newcommand{\ket}[1]{\left|#1\right\rangle}
\newcommand{\braket}[2]{\left\langle#1 |  #2\right\rangle}
\newcommand{\be}{\begin{equation}}
\newcommand{\ee}{\end{equation}}
\newcommand{\bea}{\begin{eqnarray}}
\newcommand{\eea}{\end{eqnarray}}
\newcommand{\barr}{\begin{array}}
\newcommand{\earr}{\end{array}}
\newcommand{\bl}{\left(}
\newcommand{\br}{\right)}

\newcommand{\bls}{\left[}
\newcommand{\brs}{\right]}
\def\te{\text}

\def\beq{\begin{equation}}
\def\eeq{\end{equation}}
\def\be{\begin{equation}}
\def\ee{\end{equation}}
\def\bea{\begin{eqnarray}}
\def\eea{\end{eqnarray}}

\newcommand{\eq}[1]{Eq.~(\ref{#1})}

\title{Decoherence delays false vacuum decay}
\author{Thomas C. Bachlechner}

\begin{document}
\begin{titlepage}
\baselineskip=15.5pt
\pagestyle{plain}
\setcounter{page}{1}

\bigskip\
\begin{center}
{\Large \bf Decoherence delays false vacuum decay}
\vskip 15pt
\end{center}
\vspace{0.5cm}
\begin{center}
{\large Thomas C. Bachlechner}

\vspace{0.2cm}

\vskip 4pt
\textsl{Department of Physics, Cornell University,
Ithaca, NY USA 14853}\\
\vskip 4pt
\end{center}
\small  \noindent  \\[0.2cm]
\noindent
\vspace{1cm}
\hrule \ \\
\noindent {\bf Abstract} \\
We show that gravitational interactions between massless thermal modes and a nucleating Coleman-de\,Luccia bubble may lead to efficient decoherence and strongly suppress metastable vacuum decay for bubbles that are small compared to the Hubble radius. The vacuum decay rate including gravity and thermal photon interactions has the exponential scaling $\Gamma\sim\Gamma_{\te{CDL}}^{2}$, where $\Gamma_{\te{CDL}}$ is the Coleman-de\,Luccia decay rate neglecting photon interactions. For the lowest metastable initial state an efficient quantum Zeno effect occurs due to thermal radiation of temperatures as low as the de Sitter temperature. This strong decoherence effect is a consequence of gravitational interactions with light external mode. We argue that efficient decoherence does not occur for the case of Hawking-Moss decay. This observation is consistent with requirements set by Poincar\'e recurrence in de Sitter space.

\bigskip
\hrule
\vfill \today
\end{titlepage}
\newpage

\tableofcontents

\begin{center}
\line(1,0){350}
\end{center}
\section{Introduction}

The decay of metastable vacua has been extensively studied and plays a central role in a broad class of cosmological models. The tunneling rate of a single scalar field at a metastable minimum is determined by the bounce solution of the Euclidean equation of motion, as originally demonstrated by Coleman in Ref.\,\cite{Coleman:1977py}. Effects due to coupling to gravity were considered in Ref.\,\cite{Coleman:1980aw}. However, in a de Sitter universe there are thermal gravitational modes and realistic cosmologically models have other fields that interact with the tunneling field at least gravitationally. Even though these couplings are Planck suppressed, environmental modes can lead to efficient decoherence, and thus strongly affect the dynamics of a quantum tunneling process.

In this paper we study false vacuum decay, including gravitational couplings to de Sitter modes, considering the specific example of de Sitter photons. Our goal is to determine if the decoherence induced by these interactions is sufficient to significantly suppress the tunneling rate. We find that even though the coupling is Planck suppressed and the wavelength of the de Sitter modes is of order the Hubble radius, decoherence has a significant effect on the vacuum decay rate for vacua that slowly decay via Coleman-de\,Luccia (CDL) instantons. The decoherence effect can be modeled as a quantum Zeno effect in which the wave function of the tunneling field ``collapses'' to a classical configuration each time the background leaks information to the environment about whether a bubble exists or not.

Previous works have considered decoherence from modes that are excited by the tunneling field (see e.g.\ Ref.\,\cite{Kiefer:2010pb,Queisser:2010ee,Matsumoto:2010tg}), taking into account the full master equation that governs the time evolution of the nucleating bubble and all interactions. In this work we restrict ourselves to external modes, so that we can use an S-matrix approach to evaluate the decoherence. This allows us to model the interaction as an ideal partial measurement and greatly reduces the complexity of the problem while keeping a fairly generic form of the interaction. We demonstrate that decoherence due to external modes is far more efficient than decoherence due to modes that are excited by the tunneling field.

The organization of this paper is as follows. In \S\ref{qmdecoherence} we briefly review how decoherence leads to a delay in the time evolution of a quantum system. Next,  in \S\ref{functional} we carefully demonstrate how and under which conditions a field tunneling between two minima in a quantum field theory can be described effectively by a quantum mechanical two-level system using the functional Schr{\"o}dinger method. We use these results in \S\ref{vacuumdecay} to determine how decoherence from de Sitter photons influences the bubble nucleation rate. In \S\ref{recurrence} we remark on the differences between Coleman-de\,Luccia (CDL) instantons and Hawking-Moss (HM) decay regarding decoherence, and explain how these differences ensure that de Sitter vacua do not survive longer than the recurrence time. We conclude in \S\ref{conclusion}.

\section{Decoherence and the Quantum Zeno Effect}\label{qmdecoherence}

Let us consider a simple measurement experiment in which a detector is used to determine the state of some two-level system (see e.g.\ Ref.\,\cite{Zurek:2003,Schlosshauer:2003zy,schlosshauer}). Initially, the detector and the system are uncorrelated: $\ket{\psi}= \ket{\psi_{\text{in}}}_{\text{det}} \otimes \ket{\psi}_{\te{sys}}$. Suppose that the interaction Hamiltonian is aligned with the basis $\{\ket{\uparrow}_{\te{sys}},~\ket{\downarrow}_{\te{sys}}\}$. Then after some time we can write
\begin{eqnarray}
\ket{\uparrow}_{\text{sys}}\ket{\psi_{\text{in}}}_{\text{det}}& \rightarrow &\ket{\uparrow}_{\text{sys}} \ket{\psi_{\uparrow}}_{\text{det}}\label{3.1}\\
\ket{\downarrow}_{\text{sys}}\ket{\psi_{\text{in}}}_{\text{det}} &\rightarrow &\ket{\downarrow}_{\text{sys}} \ket{\psi_{\downarrow}}_{\text{det}}\label{3.2}.
\end{eqnarray}
Here, we simply relabeled the detector state according to the state it measures. If the two-level system initially is in a coherent superposition $(\ket{\uparrow}_{\text{sys}}+ \ket{\downarrow}_{\text{sys}})/\sqrt{2}$, the state of the full system is given by
\be
{1\over \sqrt{2}}\left(\ket{\uparrow}_{\text{sys}}\ket{\psi_{\uparrow}}_{\text{det}}+ \ket{\downarrow}_{\text{sys}}\ket{\psi_{\downarrow}}_{\text{det}}\right)\, ,
\ee
and we find the reduced density matrix of the measured system by tracing over the detector:
\bea
\hat{\rho}_{\text{sys}}
={1\over2}\begin{pmatrix} 1 & \braket{\psi_{\downarrow}}{\psi_{\uparrow}}_{\te{det}} \\ \braket{\psi_{\uparrow}}{\psi_{\downarrow}} _{\te{det}} & 1 \end{pmatrix}.
\eea
Recalling that the off-diagonal entries parametrize the amount of coherence, we immediately see that for $\braket{\psi_{\uparrow}}{\psi_{\downarrow}}_{\te{det}}=0$, all coherence is lost, and the system is reduced to a classical mixture of the two basis states. This matches the intuitive result: once the detector has uniquely determined the state of the system (which corresponds to $|\braket{\psi_{\uparrow}}{\psi_{\downarrow}}_{\te{det}}|=0$) the wave function ``collapses'' to one of the eigenstates of the interaction Hamiltonian. To quantify the degree of decoherence that occurs we define the decoherence factor $r$ as
\be
r=\braket{\psi_{\uparrow}}{\psi_{\downarrow}}_{\te{det}}.\label{decoherence}
\ee
Note that at no point did we make reference to the size of the detector. It is possible to destroy all coherence of a system if it gets permanently entangled with a single quantum object. In particular, if the detector is entangled with the system and immediately brought out of causal contact we can be certain that the system has lost all coherence. This intuitive observation will turn out to provide a simple mechanism for decoherence in the case of Coleman-de\,Luccia bubble nucleation.

To see how a quantum Zeno effect arises from interaction with a single quantum object, consider a two-level system that evolves from the state $\ket{\Psi_{1}}_{\te{sys}}$ to the state $\ket{\Psi_{2}}_{\te{sys}}$ via quantum tunneling. This central system interacts with an environment that is initially uncorrelated. For $t\ll 1/\Gamma$, where $\Gamma$ is the transition rate, this system can be described by the Hamiltonian
\be
\hat{H}=\epsilon \hat{\sigma}_{z}^{\text{sys}}+\Gamma \hat{\sigma}_{x}^{\text{sys}} +\hat{H}^{\te{env}}+\hat{H}^{\te{int}}\, ,
\ee
where $\sigma_{i}$ are the usual Pauli matrices defined in the $\{\ket{\uparrow}$, $\ket{\downarrow}\}$ basis as
\be
\sigma_{x}=\ket{\uparrow}\bra{\downarrow}+\ket{\downarrow}\bra{\uparrow},~\sigma_{y}=-i\ket{\uparrow}\bra{\downarrow}+i \ket{\downarrow}\bra{\uparrow},~\sigma_{z}=\ket{\uparrow}\bra{\uparrow}-\ket{\downarrow}\bra{\downarrow}.
\ee
Furthermore, we assume that $\ket{\Psi_{1}}_{\te{sys}}$ and $\ket{\Psi_{2}}_{\te{sys}}$ are eigenstates of the interaction Hamiltonian, i.e. this is the preferred basis of the environment and we can write the interaction Hamiltonian\footnote{For simplicity we choose our basis such that $\ket{\Psi_{1}}_{\te{sys}}$ and $\ket{\Psi_{2}}_{\te{sys}}$ are eigenstates of $\sigma_{z}^{\te{sys}}$ with opposite eigenvalues.} as $\hat{H}^{\te{int}}=\hat{B}^{\te{env}}\sigma_{z}^{\te{sys}}$. This is equivalent to the statement that the environment is sensitive to whether the system is in the $\ket{\Psi_{1}}_{\te{sys}}$ or $\ket{\Psi_{2}}_{\te{sys}}$ state. We are interested in the decay probability, e.g.\ the probability for the central system to transition between its two eigenstates after interaction with the environment. Ignoring interactions, one immediately sees that the decay probability for the above Hamiltonian is given by $P_{\te{decay}}(t)=\sin^{2}(\Gamma t)\approx \Gamma^{2}t^{2}$, where $t\ll1/\Gamma$ is used in the last approximation. 

To be concrete, let the central system initially be in the state $\ket{\Psi_{1}}_{\te{sys}}$. Thus, the decay probability is given by $P_{\te{decay}}={(1-\langle \hat{\sigma}_{z}^{\text{sys}}\rangle)/2}$. The time evolution of $\langle \hat{\sigma}_{z}^{\text{sys}}\rangle$ is then
\begin{align}{d\langle\hat{\sigma}_{z}^{\text{sys}}\rangle\over dt}=i\langle[\hat{H},\hat{\sigma}_{z}^{\text{sys}}]\rangle+\left\langle {\partial\hat{\sigma}_{z}^{\text{sys}}\over \partial t}\right\rangle =2\Gamma \langle  \hat{\sigma}_{y}^{\text{sys}}\rangle\label{eh}.
\end{align} 
Considering the intrinsic evolution of the full system\footnote{Here, $\ket{\psi_{\Psi_{1,2}}}_{\te{env}}$ is the time evolution of the environment when the system is in the state $\ket{\Psi_{1,2}}_{\te{sys}}$.}, $\ket{\psi(t)}^{0}=\ket{\Psi_{1}}_{\te{sys}}\ket{\psi_{\Psi_{1}}}_{\te{env}}-i\Gamma t\ket{\Psi_{2}}_{\te{sys}}\ket{\psi_{\Psi_{2}}}_{\te{env}}+\mathcal{O}(\Gamma^{2}t^{2})$, we get
\be {d\langle\hat{\sigma}_{z}^{\text{sys}}\rangle\over dt} \approx -4\Gamma^{2} t ~\te{Re}[r(t)],~ r(t)=\braket{\psi_{\Psi_{1}}}{\psi_{\Psi_{2}}}_{\te{env}}\, .
\ee
Thus, for short times the decay probability is given by
\be
P_{\te{decay}}(t)=2\Gamma^{2}\int_{0}^{t}dt'~t'~\te{Re}\left[r(t')\right] +\mathcal{O}(\Gamma^{4}t^{4})\label{sur2}.
\ee


For $r(t)=1$, the short-time behavior of the isolated system is reproduced. It follows from Eq.~(\ref{sur2}) that as the decoherence factor approaches zero, the tunneling probability stops increasing. The source of this damping, however, is not immediately obvious. The tunneling rate can be affected when the environment is arranged in such a way that the energy levels of the central system are shifted. Then, the decoherence factor changes by a phase $e^{i\phi(t)}$, and the tunneling probability is affected even though the central system does not get entangled with the environment (e.g.\ the environment may consist of one-level systems). However, when an environment is considered that interacts but does not shift the energy levels, the central system leaks information about its state and gets entangled with the environment, such that the absolute value of the decoherence factor decreases. These two processes, which change the survival probability, are complementary.

Note that at no point did we have to make reference to the full master equation for the reduced density matrix that includes the backreaction due to the intrinsic time evolution. This is because we took the preferred basis of the interaction to be aligned with the states between which the central system transitions, i.e. $[\hat{H}^{\te{int}},\hat{\sigma}_{z}^{\te{sys}}]=0$, and because the interaction lasts only for timescales over which the intrinsic dynamics of the system can be neglected. Let us consider a decoherence factor that decays exponentially with time, say $r\sim e^{-\Gamma_{\te{dec}}t}$, which resembles repeated ideal measurements with period $1/\Gamma_{\te{dec}}$. In particular, repeated ideal measurements can be described by an S-matrix approach where a detector ``scatters'' off the system. While these are strong assumptions that do not hold for many scenarios considered in the previous literature (see Ref.\,\cite{Kiefer:2010pb, Queisser:2010ee}), it will turn out that they are satisfied for the interactions considered in this work, namely, gravitational interactions of a true vacuum bubble with massless de Sitter modes.

\section{Functional Schr\"{o}dinger Method and Metastable Vacuum Decay}\label{functional}
In the previous section we observed how decoherence may lead to suppression of a quantum tunneling process via interactions with the environment\footnote{Possible implications of decoherence in cosmology were considered in e.g.\ Ref.\,\cite{Winitzki:2007cf,Krauss:2008pt}.}. To use the same tools to study bubble nucleation we now carefully match the field theory problem of bubble nucleation to an equivalent quantum mechanics problem.

In the following, we will review the functional Schr\"{o}dinger method which we will use to derive an effective Hamiltonian that governs the quantum mechanics of the nucleating bubble. The scalar field theory we consider has an $O(4)$-symmetric solution after rotating to Euclidean space. Thus, the instanton solution can be parametrized by one variable, $\lambda$. Once the bubble solution $\phi(\lambda)$ is found, we are interested in how long it will take for the system to tunnel from the metastable vacuum to a field configuration from which the bubble solution can evolve classically. Considering the lowest metastable initial state, for times $\tau \ll \tau_{\te{CDL}}$ we can approximate the system as a two-level system in quantum mechanics. The effective two-level system has a transition time $\tau_{\te{CDL}}$ which needs to be carefully evaluated as a quantum Zeno effect only occurs for decoherence times $\tau_{\te{dec}}\ll \tau_{\te{CDL}}$. Once we obtain the effective Hamiltonian for the intrinsic time evolution of the bubble, we turn to determining the coupling to thermal de Sitter photons. The interaction between the bubble and photons can be treated in an S-matrix approach using the gravitational cross section of a bubble of critical size, which is readily available.

\subsection{Functional Schr\"odingier approach}
We first examine how the field theory problem can be mapped to a quantum mechanical system (we closely follow Ref.\,\cite{Tye:2009rb,Bitar:1978vx}).
Consider the scalar field theory
\be
{\mathcal L}= {1\over 2} \partial_{\mu}\phi \partial^{\mu}\phi-V(\phi)\, ,
\ee
where $V(\phi)$ can be any potential. For concreteness we consider the special case of the double well potential
\be
V(\phi)={g\over 4} (\phi^{2}-c^{2})^{2}- B(\phi+c)\, .\label{potential}
\ee
There exists a false vacuum at $\phi=-c$ and a true vacuum at $\phi=c$. The energy difference between the two vacua is approximately $\epsilon\approx 2 B c$. The general idea is the following: First, we map the field theory problem to an equivalent quantum mechanical tunneling problem in one dimension. Expanding around the false vacuum solution provides us with the metastable ground state solution that will tunnel through the effective potential describing the full double well in field theory. In the vicinity of the metastable vacuum $\phi=-c$ the potential is given by
\be\label{false}
V_{\te{cl}}= c^2 g (\phi + c)^2-B(\phi+c)+{\mathcal O}(\phi^{3})\, .
\ee
The theory is quantized by demanding the relation $[\dot{\phi}({\bf x}),\phi({\bf x^{\prime}})]=-i \hbar \delta^{3}({\bf x}-{\bf x^{\prime}})$. The resulting functional Hamiltonian is given by
\be
H=
\int d^{3}{\bf x}\bl -{\hbar ^{2}\over2} \bl{\delta\over \delta \phi({\bf x})}\br^{2}+{(\nabla \phi)^{2}\over2}+V(\phi) \br.\label{eq:Hamiltonian}
\ee
Considering the Hamiltonian (\ref{eq:Hamiltonian}) we can define an effective potential
\be
U(\phi)=\int d^{3}{\bf x}\bl {1\over 2} (\nabla \phi)^{2}+V(\phi)\br\label{effectivepot}.
\ee

Using the ansatz $\Psi(\phi({\bf x}))= A \exp(-i S(\phi({\bf x}))/\hbar)$ and expanding in powers of $\hbar$, such that $S=S_{0}(\phi)+\hbar S_{1}(\phi)+\dots$, we can write the functional Schr\"{o}dinger equation at leading and next to leading order
\bea
&&\int d^{3}{\bf x} \bls {1\over 2} \bl{\delta S_{0}(\phi)\over \delta \phi}\br^{2} +{1\over 2} (\nabla \phi )^{2}+V(\phi)\brs =E\, ,\label{eq:leading}\\
&&\int d^{3}{\bf x} \bls - {i} {\delta^{2} S_{0}(\phi)\over \delta \phi^{2}} +2{\delta S_{0}\over \delta\phi} {\delta S_{1}\over \delta\phi} \brs =0\, .\nonumber
\eea
We are interested in the most probable escape path (MPEP), that is, the path $\phi({\bf x},\lambda)$ that continuously interpolates between the false and the true vacuum as the parameter $\lambda$ is varied such that the action is minimized in the transverse directions. Let $\phi({\bf x},\lambda)$ be a path and define a length along this path in field space as $ds^{2}=\int d^{3}{\bf x}~[d\phi({\bf x},\lambda)]^{2}$. We can also write this length in terms of $d\lambda$ as
\be
ds=\left({\int d^{3}{\bf x}~\left[{\partial\phi({\bf x},\lambda)\over \partial \lambda}\right]^{2}}\right)^{1/2} d\lambda\, .
\ee
We can define a tangent vector along the path as
\be
\delta \phi_{\parallel}={\partial \phi\over \partial s} ds,
\ee
and a perpendicular vector
\be
\delta \phi_{\perp}=\delta \phi-a ds {\partial \phi\over \partial s}\, ,
\ee
with $a$ such that
\be
\int d^{3} {\bf x}~ \delta\phi_{\perp}{\partial \phi\over \partial s}=0\, .
\ee
The most probable escape path in $\phi$ space is chosen such that the variation of $S_{0}$ vanishes in the perpendicular direction, while the variation does not vanish in the parallel direction. We can parametrize the MPEP by $\lambda$, which leads to \cite{Bitar:1978vx}
\bea
&&{\delta S_{0}\over \delta\phi_{\parallel}}\bigg|_{\phi_{0}}={\partial S_{0}\over \partial \lambda}\left( \int d^{3}{\bf x}\left[ {\partial\phi\over \partial \lambda }\right]^{2} \right)^{-1} {\delta \phi_{0}\over \delta \lambda}\\ &&{\delta S_{0}\over \delta\phi_{\perp}}\bigg|_{\phi_{0}}=0\, .
\eea
In Ref.~\cite{Bitar:1978vx} it is demonstrated how to solve the WKB equations \label{eq:leading} at leading order along the MPEP which determines $\phi({\bf x},\lambda)$. The Euler-Lagrange equation for $\phi$ becomes in the classically forbidden region $U(\phi)>E$
\be
{\partial^{2} \phi({\bf x},\tau) \over \partial \tau^{2}}+\nabla^{2} \phi({\bf x},\tau)-{\partial V( \phi({\bf x},\tau))\over \partial \phi}=0\, ,\label{forbidden}
\ee
where $\tau$ is the Euclidean time can be related to the variable $\lambda$ parametrizing the MPEP. Eq.~(\ref{forbidden}) allows the $O(4)$ symmetric domain wall solution (in the thin-wall approximation)
\be\label{phi0}
 \phi({\bf x},\lambda)=-c~ \te{tanh}\bl  {\mu\over 2} (\sqrt{\tau^{2}+|{\bf x}|^{2}}-\lambda_{c}) \br\approx -c ~\te{tanh}\bl  {\mu\over 2} {(|{\bf x}|^{2}-\lambda^{2})\over 2 \lambda_{c}} \br\,
\ee
where $\mu=\sqrt{2g c^{2}}$, $\lambda=\sqrt{\lambda_{c}^{2}-\tau^{2}}$, and $\lambda_{c}$ is determined by considering the balance between the domain wall tension $S_{1}$ and the vacuum energy:
\be
S_{E}=-{\pi^{2}\over 2} \lambda^{4}+2\pi^{2}\lambda^{3} S_{1},
\ee
with the domain wall tension 
\be
S_{1}=\int_{-c}^{c}d\phi \sqrt{2V(\phi)}\approx \sqrt{g\over 2} {4 c^{3}\over 3}.
\ee
Setting the variation of the total action to zero we find the critical radius of the bubble $\lambda_{c}=3S_{1}/\epsilon$. Any bubble smaller than $\lambda_{c}$ will decay while any bubble larger than $\lambda_{c}$ will grow classically.

\begin{figure}[h]
	\centering
		\includegraphics[width=0.5\textwidth]{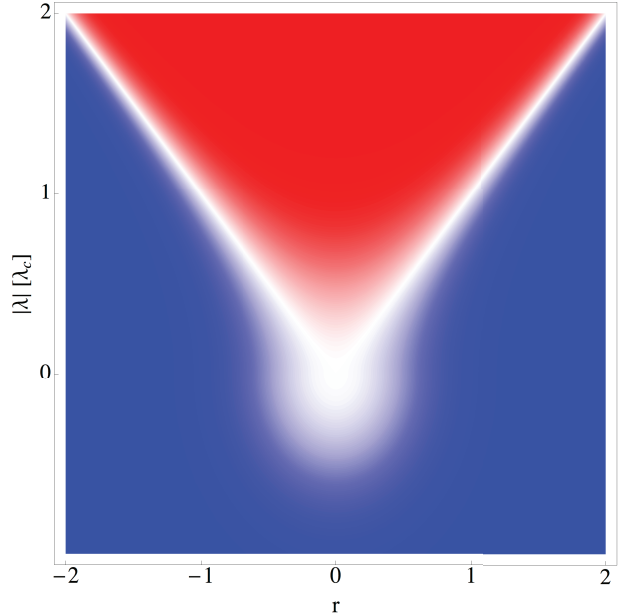}
	\caption{\small Contour plot of $\phi(r,\lambda)$, where $\lambda^{2}$ varies from negative to positive values. Red corresponds to the true vacuum $\phi=c$ while blue corresponds to the false vacuum $\phi=-c$.}
\label{bubble}
\end{figure}

In the classically allowed region, the solution to the Euler-Lagrange equation is given by Eq.~(\ref{phi0}) with $\lambda=\sqrt{\lambda_{c}^{2}+\tau^{2} }$. Note that in order to have a continuous parameter that describes the MPEP we are required to have $\lambda^{2}$ vary continuously from negative to positive values. Thus, in the following we choose $\lambda^{2}$ as parametrizing the MPEP. To illustrate the nucleation and expansion of a bubble via the MPEP, Figure~\ref{bubble} shows $\phi({\bf x},\lambda^{2})$ over $|\lambda|$ where, again, $\lambda^{2}$ varies from negative to positive values in order to capture both the classically allowed and forbidden regions. 

Now that we obtained an explicit approximation for the most probable escape path, we can consider the quantum mechanical problem of tunneling from the false vacuum to the true vacuum. Substituting Eq.~(\ref{phi0}) in the Hamiltonian (\ref{eq:Hamiltonian}) gives
\bea\label{hamilt}
H[\phi({\bf x},\lambda^{2})]
&=&{p_{\lambda^{2}}\over 2m(\lambda^{2})}+U(\phi({\bf x},\lambda^{2})\, ,
\eea
where we defined a $\lambda$ dependent mass
\be
m(\lambda^{2})=\int d^{3}{\bf x} ~ \left({\partial \phi\over \partial \lambda^{2}}\right)^{2} \, ,
\ee
and the momentum is given by
\be\label{momentum}
p_{\lambda^{2}}=m(\lambda^{2})  \dot{\lambda^{2}}\, .
\ee
Here, we can interpret the variable $\lambda^{2}$ as an effective position along which the wave functional $\Psi$ evolves. Combining (\ref{hamilt}) with (\ref{momentum}) we find the quantum mechanical Hamiltonian as
\be
H\Psi(\lambda^{2})=\left[-{1\over 2m(\lambda^{2}) }\left({\partial\over \partial \lambda^{2}}\right)^{2}+U(\lambda^{2})\right]\Psi(\lambda^{2})\, . \label{eq:h}
\ee

\begin{figure}[h]
	\centering
		\includegraphics[width=0.5\textwidth]{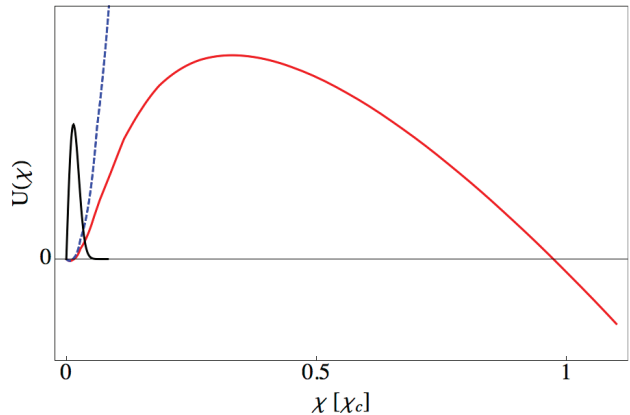}
	\caption{\small The wavefunction of the lowest metastable initial state (black line) is shown along with the classical potential (dashed, blue line) and the full effective potential (red line) over the rescaled variable $\chi$.}
\label{qmpicture}
\end{figure}

In order to estimate the tunneling probability we can use the WKB approximation and the effective potential in \eq{effectivepot} to obtain the solution to the functional Schr\"{o}dinger equation. For a bubble at critical radius $\lambda_{c}$ one obtains \cite{Tye:2009rb}
\be
\Psi(\phi({\bf x}, \lambda_{c}))=A \exp\bl-{ 1\over\hbar}\int_{0}^{\lambda_{c}}d\lambda \sqrt{2m(\lambda) [U(\lambda)-E]}\br\sim A \exp\bl-{ \pi^{2}\over4\hbar}S_{1}\lambda_{c}^{3}\br\, .\label{wkbsol}
\ee
Thus, the tunneling rate can be written as
\be
\Gamma_{\te{CDL}}\sim |A|^{2} \exp\bl-{ \pi^{2}\over2\hbar}S_{1}\lambda_{c}^{3}\br\, ,\label{cdlrate}
\ee
which is precisely the Coleman-de Luccia vacuum decay rate. This result deserves some discussion. First, note that while \eq{wkbsol} is just the same exponential scaling as found in Ref.\,\cite{Coleman:1980aw}, we only solved a time-independent one dimensional quantum mechanics problem\footnote{Note that the exponential in \eq{wkbsol} differs from the result for the tunneling rate in Ref.\,\cite{Coleman:1980aw} by a factor of two. This is because we calculated the tunneling amplitude rather than the tunneling rate.}. However the present position dependent mass obstructs some of the intuition from standard quantum mechanics. In particular, the potential vanishes approximately for $\lambda^{2}<0$ as this corresponds to the homogeneous false vacuum solution so it is not clear how to define an initial state in this potential. In order to avoid the position dependence of the mass we transform to a new coordinate that absorbs the position dependence. Let
\be\label{chidef}
{d\chi \over d\lambda^{2}}=\sqrt{m(\lambda^{2})}\, .
\ee
With the new variable $\chi$ in (\ref{chidef}) the Hamiltonian (\ref{eq:h}) becomes
\be
H=-{1\over 2}\left({\partial\over \partial \chi}\right)^{2}+U(\lambda^{2}(\chi))\, .
\ee
Note that $m(\lambda^{2})\approx 0$ for $\lambda^{2}<0$ such that $\lambda^{2}=-\infty$ can be mapped to $\chi=0$. This is a very useful identification as it allows to localize the wave function corresponding to the false vacuum solution at finite $\chi$. 
Using the potential (\ref{potential}) and the most probable escape path (\ref{phi0}) we can evaluate the mass and potential in terms of $\lambda^{2}$:
\bea
U(\lambda^{2})&\approx &{4\pi c^2  \sqrt{\lambda^{2}} \mu\over 3\lambda_{c}}(\lambda_{c}^2- \lambda^2) \\
m(\lambda^{2})&\approx& {2\pi c^{2}\sqrt{\lambda^{2}}\mu\over 3\lambda_{c}}\nonumber\, .
\eea
Using (\ref{chidef}) we can rewrite these expressions in terms of the rescaled variable $\chi$
\be
\chi(\lambda^{2})\approx \frac{4}{5} \sqrt{\frac{2 \pi\mu c^{2} }{3\lambda_{c}}} (\lambda^{2})^{5/4}\, .
\ee
For the potential this gives with $\chi_{c}=\chi(\lambda_{c}^{2})$
\be
U(\chi)\approx  {5 \sqrt{\pi \mu c^{2}} \over \sqrt{6}} \left(\frac{\chi}{\chi_{c}}\right)^{2/5} \left(\chi_{c}-\left(\chi ^4\chi_c\right)^{1/5}\right)\, .
\ee
As mentioned before, the initial metastable vacuum state is given by the ground state of the effective potential expanded around the false vacuum $\phi=-c$. To evaluate the initial state wavefunction we require the effective potential from the expansion around the false vacuum $V_{\te{cl}}$ in (\ref{false})
\be
U_{\te{cl}}(\chi)\approx \left({2\pi 5^{6} \over 3\mu^{3}}\right)^{1/4}\sqrt{g^{2}c^{5}\chi_{c}} \left({\chi^{6}\over \chi_{c}}\right)^{1/5} \, .
\ee
Now, we fully reduced the tunneling problem to a quantum mechanical problem in one dimension with constant mass. Solving for the ground state in the false vacuum effective potential gives the lowest metastable initial state. Subsequently, this state is placed in the full effective potential $U(\chi)$ that allows for tunneling. The initial metastable state can be approximated as the superposition of two energy eigenstates that are separated by approximately $\Delta E=\Gamma_{\te{CDL}}$. To illustrate this scenario, the metastable wave function is evaluated numerically and shown in Figure \ref{qmpicture} along with the classical and full potentials.

 A possible concern is that interference effects from bubbles of different radii or bubbles at other positions alter the tunneling dynamics. The tunneling rate decreases exponentially with bubble radius. If we are interested in the state of the system at times of order $1/\Gamma(\lambda_{c})$, bubbles of smaller radius will have vanished, while bubbles of larger radius have an exponentially suppressed amplitude. Furthermore, for $H\gg\Gamma(\lambda_{c})$ only one classically expanding bubble is nucleated per Hubble volume, so that interference effects from other bubbles can be neglected consistently. Of course, this is only true for potentials that do not allow for resonant tunneling, in which case the situation may become more complicated (see e.g.\ Ref.\,\cite{Tye:2009rb}). 
\subsection{Approximate two-level system}\label{two-level}
As argued above we are interested in modeling the time evolution of the tunneling process as an approximate two-level system. This system evolves from the homogeneous false vacuum solution to a bubble of critical size. For times $t\ll 1/\Gamma_{\te{eff}}$ we can define the effective Hamiltonian\footnote{We choose our basis such that $\hat{\sigma}_{z}\Psi_{\te{false}}=+\Psi_{\te{false}}$ and $\hat{\sigma}_{z}\Psi_{\te{true}}=-\Psi_{\te{true}}$.} (see also Ref.\,\cite{Kiefer:2010pb,Queisser:2010ee})
\be\label{Hamiltonian}
\hat{H}_{\te{Bubble}}={2\pi\over 3}R_{c}^{3} \Bigl(V(\phi_{\te{true}}) (1+\hat{\sigma}_{z})+V(\phi_{\te{false}})(\hat{\sigma}_{z}-1)\Bigr)+ \Gamma_{\te{eff}} \hat{\sigma}_{x}\,,
\ee
where $\Gamma_{\te{eff}}$ is an effective decay rate that depends on the energy spectrum of the metastable initial state. The description of the bubble nucleation process as a two-level system relies on the assumption that the non-decay probability decreases quadratically as $P_{\te{non-decay}}\approx 1-\Gamma_{\te{eff}}^{2}t^{2}+\mathcal{O}(\Gamma_{\te{eff}}^{4}t^{4})$. Note also that $\Gamma_{\te{eff}}\sim \Delta E_{\te{max}}$, where $\Delta E_{\te{max}}$ is the largest energy difference contained in the energy spectrum of the initial state. To make this statement more precise, note that we can write the non-decay probability in terms of the energy spectrum of the initial metastable state $\sigma(\mu)$ as $P(t)=|a(t)|^2$, where (see e.g. Ref.~\cite{Chiu:1977ds})
\be
a(t)=\int d\mu~ \sigma(\mu) e^{-i \mu t}~,~
\sigma(\mu)=|\braket{\phi_{\mu}}{\psi_{0}}|^{2}.\label{nondecay}
\ee
Here, $\ket{\phi_{\mu}}$ are the eigenstates of the Hamiltonian. From \eq{nondecay} we see that the non-decay probability is constant at least for times on the order of $1/\Delta E_{\te{max}}$. The relevant quantity that determines the non-decay probability and thus the time for which the system can be modeled as an approximate two level system is the energy spectrum $\sigma(\mu)$ of the initial state. In general, the spectrum needs to be computed for a specific initial state which leads to some effective decay rate $\Gamma_{\te{eff}}$. In this work we constrain ourself to the lowest metastable initial state, i.e. the lowest energy eigenstate of the potential expanded around the false vacuum. We numerically verified that the spectrum of the lowest metastable initial state has a Gaussian spectrum such that it can be modeled as an approximate two level system with an effective decay rate $\Gamma_{\te{eff}}\sim \Gamma_{\te{CDL}}$. When considering an excited initial state that is not the lowest metastable false vacuum state, the effective decay time $\tau=1/\Gamma_{\te{eff}}$ may be small compared to $\tau_{\te{CDL}}$ (which was computed in the zero energy approximation). As we do not attempt any quantitative analysis but rather illustrate the mechanism of decoherence we do not consider any excited initial states\footnote{The qualitative results of this work remain valid for an arbitrary initial state but will require the stronger bound $\tau_{\te{dec}}\ll 1/\Gamma_{\te{eff}}$.}.

Now that we have established that false vacuum decay can be modeled by a two-level system with intrinsic Hamiltonian (\ref{Hamiltonian}), where $\Gamma_{\te{eff}}=\Gamma_{\te{CDL}}$, we are in a position to consider additional contributions to the Hamiltonian. Any realistic cosmological model allows for fields other than just one isolated scalar. To capture possible effects on tunneling due to environmental degrees of freedom we write the full Hamiltonian in the schematic form
\be
\hat{H}=\hat{H}_{\te{Bubble}}+\hat{H}_{\mathcal{E}}+\hat{H}_{\te{int}},
\ee
where all fields other than $\phi$ are absorbed in the environmental part $\hat{H}_{\mathcal{E}}$. Note that by modeling the bubble as an effective two-level system and neglecting the classical growth after nucleation we underestimate the bubble-photon coupling, and thus obtain a lower bound on the environment induced decoherence. 

\section{Decoherence and False Vacuum Decay}\label{vacuumdecay}
The conclusions of the previous two sections apply for the lowest metastable initial state and generic bubble-environment interactions that can be modeled by an S-matrix approach, i.e.\ external modes that interact with the nucleating bubble for a short time during which the intrinsic bubble evolution is negligible. We now turn to a specific environment, consisting of de Sitter photons coupled to gravity, to obtain the decoherence rate and demonstrate the emergence of an efficient quantum Zeno effect. This is a minimalistic approach towards decoherence to demonstrate the mechanism. In a generic setup there will be other massless excitations that lead to far stronger decoherence effects than those due to de Sitter photons. On the other hand, an excited initial state may decrease the effective decay time $1/\Gamma_{\te{eff}}$ and requires careful treatment.

\subsection{Particle Interaction}
Consider a nucleating bubble $\ket{{\bf x}}$ at position ${\bf x}$, coupled to an environment of modes $\ket{\chi}_{i}$ where the interaction is well described by an S-matrix approach (see Ref.\,\cite{schlosshauer} for a detailed discussion). Initially, the environment and the bubble are uncorrelated, so the full density matrix factorizes as
\be
\hat{\rho} (0)=\hat{\rho}_{\te{B}}(0)\times \hat{\rho}_{\mathcal{E}}(0).
\ee
We are evaluating the decoherence factor in position space: $r({\bf x},{\bf x^{\prime}},t)$. This is just the quantity we are interested in, as when coherence over a distance $|{\bf x}-{\bf x^{\prime}}|=\lambda_{\te{c}}$ is lost, the MPEP is inaccessible and the bubble nucleation process is highly suppressed. Remember that the decoherence factor is the off-diagonal element of the reduced density matrix, which is given by
\be
\hat{\rho}_{\te{B}}=\te{Tr}_{\mathcal{E}}\hat{\rho}=\int d{\bf x}d{\bf x^{\prime}}~ \rho_{\te{B}}({\bf x},{\bf x^{\prime}},0) \ket{\bf x}\bra{\bf x^{\prime}} \braket{\chi({\bf x^{\prime}})}{\chi({\bf x})}.
\ee 
Assuming no momentum transfer, an isotropic distribution of scattering particles, and a slow intrinsic bubble evolution, the off-diagonal matrix element of the reduced density matrix is determined by (see e.g.\ Ref.\,\cite{schlosshauer})
\be
{\partial  \rho_{\te{B}}({\bf x},{\bf x^{\prime}},t)\over \partial t}=-F({\bf x}-{\bf x^{\prime}})  \rho_{\te{B}}({\bf x},{\bf x^{\prime}},t),\label{decoherenceevolution}
\ee
where
\be
F({\bf x}-{\bf x^{\prime}}) =\int dq~\nu(q)v(q) \int {d{\bf n}d{\bf n^{\prime}}\over 4\pi}\bl1-e^{i q ({\bf n}-{\bf n^{\prime}})\dot({\bf x}-{\bf x^{\prime}})} \br|f({\bf q},{\bf q'})|^{2}\,.
\ee
Here $v(q)$ is the velocity distribution, $\nu(q)$ denotes the momentum density of particles and $|f|^{2}$ is the scattering amplitude squared. In the long-wavelength limit, the off diagonal component of the density matrix is given by
\be
\rho_{\te{B}}({\bf x},{\bf x^{\prime}},t)= \rho_{\te{B}}({\bf x},{\bf x^{\prime}},0)e^{-\Lambda |{\bf x}-{\bf x^{\prime}}|^{2}t}\, ,\label{decoherencerate}
\ee
where
\be
\Lambda= {2\pi \over 3} \int dq~ \nu(q) v(q) q^{2}\bl \int d\cos(\theta)~ \bls1-\cos(\theta)\brs |f(q,\theta)|^{2}\br.
\ee
Thus, in the long wavelength limit, coherence is lost over a distance $\Delta x$ after times of order $t_{\te{dec}}\approx(\Lambda (\Delta x)^{2})^{-1}$.

\subsection{Decoherence from thermal photons}

We now use the framework of decoherence developed above to estimate the effects of interactions with thermal photons on bubble nucleation. Note that all assumptions made in Section \ref{qmdecoherence} about the interaction are satisfied for the case of gravitational scattering of photons: the interaction timescale is exponentially small compared to the vacuum decay rate and the preferred basis of the bubble-photon interaction is aligned with the true and false vacuum configuration. At this point it becomes important to check if the decoherence time is small compared to the effective vacuum decay rate, i.e. the timescale for which the bubble obeys quadratic decay and can be modeled as a two-level system. If the decoherence time is small compared to the effective decay rate we can neglect the intrinsic bubble evolution in the master equation, leading to the simple result for the decay probability found in \eq{sur2}. It will turn out that decoherence due to external modes is dominant compared to interactions with modes sourced by the tunneling field (see e.g.\ Ref.\,\cite{Kiefer:2010pb,Queisser:2010ee}).

In order to estimate the decoherence time we evaluate the cross section of gravitational bubble-photon scattering. Let us consider a static, spherically symmetric bubble of true vacuum. In the linear approximation such a configuration leads to the metric ($\eta=\te{diag}(+,-,-,-)$)
\be
g_{\mu \nu}=\eta_{\mu\nu}+\kappa h_{\mu \nu}({\bf x})=\eta_{\mu\nu}-2 \phi(r)(\eta_{\mu\nu}-2\eta_{\mu 0}\eta_{\nu 0}),\label{metric}
\ee
where $\kappa^{2}=32\pi G_{\te{N}}$ and $\phi$ is the classical potential. Once $\phi(r)$ is fixed we consider the metric to be static. The bubble interacts gravitationally with photons via the action
\be
S=-\int d^{4}x~\sqrt{g} {F_{\mu\nu}F^{\mu\nu}\over 4},
\ee
where $F_{\mu\nu}=\partial_{\nu}A_{\mu}-\partial_{\mu}A_{\nu}$. Expanding $\sqrt{g}$ around flat space gives the vertex for photon-graviton interactions (see Ref.\,\cite{accioly}):
\be
V_{\mu\nu}(p,p^{\prime}) =  {\kappa h^{\lambda \rho}({\bf k})\over 2}
\bls\eta_{\lambda\rho}p_{\nu}
p^{\prime}_{\mu}-\eta_{\mu\nu}\eta_{\lambda\rho}{\bf p}.{\bf p}^{\prime}+2\bl\eta_{\mu\nu}p_{\lambda}p^{\prime}_{\rho}-
\eta_{\nu\rho}p_{\lambda}p^{\prime}_{\mu}-
\eta_{\mu\lambda}p_{\nu}p^{\prime}_{\rho}+\eta_{\mu\lambda}\eta_{\nu\rho}{\bf p}.{\bf p}^{\prime}\br\brs,
\ee
where $h_{\mu\nu}({\bf k})=\int d^{3}{\bf x}~e^{-i {\bf k}\cdot {\bf x}}h_{\mu\nu}({\bf x})$ is the Fourier transform of $h_{\mu\nu}({\bf x})$. 
We now turn to evaluating the classical gravitational potential inside a bubble. The most probable size of a non-collapsing bubble is just the critical radius at which the surface tension is balanced by the smaller energy density inside and the gravitational energy. Assuming a bubble of critical radius, the energy in surface tension just cancels the volume energy such that the gravitational potential outside the bubble vanishes. Inside the bubble, the potential is given by $\phi(r)={\kappa^{2}}{r^{2}\epsilon}/24$.
After Fourier transforming the potential we find the polarization averaged differential cross section to be
\bea
{d\sigma\over d\Omega}&=&{1\over(4\pi)^{2}}{1\over 2}\sum_{\te{polarizations}}\left|{{\epsilon}_{r}^{\mu}}{\epsilon_{r^{\prime}}^{\nu}}V_{\mu\nu}\right|^{2}\nonumber\\
&=&{64\pi^{2}G_{\te{N}}^{2}} | I(R,k)|^{2}E^{4}(1+\cos(\theta))^{2},\label{diffcross}
\eea
where $I(R,k)=\int_{0}^{R}dr~r^{2 }e^{-i k r}  {r^{2}\epsilon/ 3}$.
Note that the cross section at photon momenta $k\ll 1/R$ scales as $\sigma\sim k^{4}$. Thus, the leading contribution to the decoherence rate is due to modes of wavelengths smaller than the Hubble radius, so that the flat space approximation we used to obtain the scattering amplitude is valid. 

\subsection{Quantum Zeno effect for the metastable ground state}
As argued in  \S\ref{two-level} we can model false vacuum decay of a metastable initial state as a two-level process using the Hamiltonian in \eq{Hamiltonian}. This is only valid for timescales $\tau\ll 1/\Gamma_{\te{eff}}$, where $\Gamma_{\te{eff}}\sim \Gamma_{\te{CDL}}$ for the lowest metastable initial state. In the following, we assume the system initially is in the metastable ground state\footnote{For more general initial states the timescale $1/\Gamma_{\te{eff}}$ may decrease which requires careful evaluation of whether the decoherence present in the model plays a significant role.}.  Furthermore, this far we only considered vacuum decay in flat space, neglecting gravity. Including gravity leads to a different effective potential and changes the critical radius above which the bubble grows classically. However, these changes can be directly translated into an equivalent quantum mechanics problem as only the effective potential changes. It is a reasonable assumption that the vacuum decay including gravity can also be modelled as an effective two level system, which we will assume in the following. Under this assumption, Eq.~(\ref{sur2}) applies also including gravity.

To obtain the decoherence rate we can combine the differential cross section in \eq{diffcross} with Eq.~(\ref{decoherencerate}). Considering de Sitter radiation at a temperature $T\ll 1/R$ we find
\bea
\Gamma_{\te{dec}}
\approx {7\pi \times 2^{16}\over 45} G_{\te{N}}^{2}\epsilon^{2} R^{12} T^{9} .\label{rate0}
\eea
The radius above which a bubble grows classically including gravity is given by $R_{\te{c}}=R_{0}/[1+( R_{\te{0}}^{2}\epsilon/(12m^{2}_{\te{pl}}))]$ (see Ref.\,\cite{Coleman:1980aw}), where $R_{0}=3S_{1}/\epsilon$ is the critical radius neglecting gravity. Substituting the critical radius in Eq.~(\ref{rate0}), the decoherence rate due to thermal de Sitter photons including gravity is given by
\be
\Gamma_{\te{dec}}\approx\frac{7\times3^{5}\times 2^{25} \sqrt{3}  S_{1}^{12}m_{\te{pl}}^{11} \epsilon^{13/2}}{5\pi^{10}  \left(3 S_{1}^{2}+4{m_{\te{pl}}^2 \epsilon }\right)^{12}},\label{dec}
\ee
where we used the de Sitter temperature $T=H/(2\pi)=\sqrt{\epsilon/(3m^{2}_{\te{pl}})}/(2\pi)$.
The decoherence rate implied by \eq{dec} is to be compared to the rate of bubble nucleation including gravity, which is given by \cite{Coleman:1980aw}
\be
\Gamma_{\te{CDL}}\approx \exp \bl- {24m_{\text{pl}}^{4}\pi^{2}S_{1}^{4}\over  \epsilon \left(1+{4\epsilon m_{\te{pl}}^{2}\over 3 S_{1}^{2}}\right)^{2}} \br\, .\label{decayrate}
\ee
Note that the approximation in \eq{decayrate} for the nucleation rate is only valid for cases where the expression inside the exponential is large, such that a polynomial prefactor, corresponding to one-loop corrections, can be neglected. The regime in which the tree level approximation for the nucleation rate is valid is just the same regime where we have $\Gamma_{\te{CDL}}\ll \Gamma_{\te{dec}}$. This is because the decoherence time is only polynomially small while the nucleation rate is exponentially small. To estimate the rate of bubble nucleation including interactions with de Sitter photons, we can combine \eq{sur2}, \eq{dec} and \eq{decayrate} and obtain
\be
\Gamma
=2\Gamma_{\te{dec}}^{-1}\Gamma_{\te{CDL}}^{2}\approx \Gamma_{\te{CDL}}^{2},
\ee
where we only kept the exponential dependence in the last approximation. This is the main result of this paper. The decoherence induced by interactions with massless external modes leads to an additional factor of $2$ in the exponent of the decay rate, indicating a strong suppression of Coleman-de\,Luccia bubble nucleation. Furthermore, even though we assumed interactions with de Sitter photons in the above example, the same qualitative features are expected from interactions with de Sitter gravitons. This is because for scattering off a classical gravitational potential, the photon cross section differs only in the angular dependence from the graviton cross section, leading to the same parametric scaling of the decoherence factor (see e.g. Ref.~\cite{Peters:1976jx}). Note that in deriving the effective nucleation rate we used the thin-wall approximation and we assumed a bubble much smaller than the de Sitter radius. 


\subsection{Comparison to previous work}
The effects of decoherence on false vacuum decay have previously been discussed in Ref.~\cite{Kiefer:2010pb,Queisser:2010ee}. In \cite{Kiefer:2010pb} a tri-linear coupling between a {\it homogeneous} tunneling field and massless environmental modes is considered, which is given by the Lagrangian
\be
{\mathcal L}={1\over 2} \dot{\phi}^{2}-V(\phi) +{1\over 2} |\partial_{\mu}\sigma|^{2}+g_{s}\sigma^{2}\phi\, .
\ee
The coupling of the tunneling field $\phi$ to an environment $\sigma$ induces decoherence that naively would lead to a suppression of the tunneling rate. However, because the field $\phi$ is assumed to be homogeneous the modes $\sigma$ never decouple from the interaction. Furthermore, in contrast to the case of a finite number of oscillators that induce decoherence, the fact that there are an infinite number of degrees of freedom requires regularisation that leads to an effect that can be interpreted analogous to the Casimir effect and causes an enhancement of the tunneling. The boundary conditions restrict the amount of decoherence that can occur and the enhancement of the tunneling rate is purely due to the fact that the $\sigma$ modes never decouple from $\phi$. An enhancement of the tunneling rate would not be expected in a quantum mechanical treatment as demonstrated in \cite{cal}, where interactions with a finite number of environmental degrees of freedom introduce an effective friction term that suppresses the time evolution.

The scenario considered in \cite{Kiefer:2010pb} is crucially different from the one considered in this work. Instead of considering excitations that are sourced by a homogeneously tunneling field $\phi$, we consider an inhomogeneous field configuration that evolves in a bath of finite temperature excitations. The homogeneous approximation is good for the case where the whole Hubble volume tunnels simultaneously and only fields that are excited by the evolution of $\phi$ can contribute to decoherence. Intuitively, this effect is weaker than external measurements as the environment continuously interacts with $\phi$. On the other hand, when external modes scatter of a tunneling system and subsequently are out of causal contact the coherence of the system is lost irreversibly at the time of the interaction. In general, treating the inhomogeneous solution $\phi$ coupled to external modes $\sigma$ in field theory is a hard problem. Instead, we reduced the situation to a quantum mechanical tunneling process of a two state system that is periodically probed by scattering with external modes. As the timescale of scattering is much smaller than the typical evolution of the background solution $\phi$ we were able to neglect the background evolution which allowed us to further simplify the problem to periodic partial measurements of an evolving two-level system which leads to a suppression of the tunneling rate.

\section{Decoherence and de Sitter Recurrence}\label{recurrence}
A possible worry is that any string theoretic description of de Sitter space becomes inconsistent at timescales larger than the recurrence time (see e.g.\ Ref.\,\cite{Kachru:2003aw} and references therein). For a single scalar field $\phi$ the timescale of CDL decay including gravity is given by $t_{\te{decay}}\sim e^{S(\phi)+{\bf S_{0}}}$, where ${\bf S_{0}}=-S(\phi_{0})=24\pi^{2}/V_{0}=\te{log}(t_{r})$ is the de Sitter entropy and $t_{r}$ is the recurrence time. Expanding (\ref{decayrate}) around small $\epsilon m_{\te{pl}}/S_{1}^{2}$ we have
\be
t_{\te{CDL}}\sim \exp \bl{24 \pi^{2 }m_{\te{pl}}^{4}\over \epsilon}-{64\pi^{2}m_{\te{pl}}^{6}\over S_{1}^{2} }\br \, .
\ee
If we consider interactions with de Sitter photons, however, we saw in Section~\ref{vacuumdecay} that the CDL decay time is changed to about $t_{\te{decay}}^{\te{dec}}\sim t_{\te{decay}}^{2}$ for certain initial states, which is at risk of exceeding the limits set by Poincar\'e recurrence. In the following we demonstrate how, despite this apparent inconsistency, the timescale of vacuum decay does not exceed the recurrence time even when interactions with photons and the resulting quantum Zeno effect are included.

There are two possible decay channels through which a false vacuum can decay. For Coleman-de\,Luccia decay a bubble of true vacuum forms that subsequently grows classically. On the other hand, for Hawking-Moss decay the whole universe tunnels homogeneously out of the false vacuum. For Hawking-Moss decay the typical timescale is given by \cite{Kachru:2003aw} 
\be
t_{\te{HM}}\sim \exp \bl{24 \pi^{2 }m_{\te{pl}}^{4}\over \epsilon}-{24 \pi^{2 }m_{\te{pl}}^{4}\over V_{1}}\br\, ,
\ee
where $V_{1}$ is the de Sitter maximum of the potential.
In Section~\ref{vacuumdecay} we demonstrated how the scattering of external modes provides an efficient mechanism for inducing decoherence. At late times we found a decoherence factor that decreases exponentially with time. This mechanism, which can only occur for CDL tunneling, is very efficient, because after the scattering the detector is out of causal contact with the system, so that coherence cannot be restored. On the other hand, if we consider HM decay in which the whole causal patch tunnels homogeneously, the S-matrix approach is not applicable anymore, as there are no external states. The scenario of continuous system-environment interaction was studied in Ref.\,\cite{Kiefer:2010pb,Queisser:2010ee}, where it was found that the decoherence factor decreases polynomially at late times, which is insufficient to induce a strong quantum Zeno effect. 

At this point it becomes important to check that a single causal patch can be treated as a closed quantum system that is independent of any physics beyond the horizon. Following Ref.\,\cite{Goheer:2002vf} let us consider the example of $3$ dimensional de Sitter space with symmetry group $SO(3,1)$. There are three rotations and three boosts. However, only one rotation (spatial rotations) and one boost (time translation) preserve the causal patch. In Ref.\,\cite{Goheer:2002vf} it is demonstrated how the other four symmetries that do not preserve the causal patch are not consistent with assigning a finite amount of entropy to a causal patch in de Sitter space, and need to be broken if the holographic principle holds. Thus, from the observer's point of view a causal patch can be treated as an isolated quantum system that does not interact with any degrees of freedom outside the horizon. This indicates that the S-matrix approach to decoherence is not applicable for the case of HM vacuum decay. Hence, we expect the decoherence factor to decrease polynomially with time such that the exponential scaling of the HM vacuum decay rate is not changed by including interactions with other degrees of freedom.

Now that we have argued that HM decay is not significantly affected by decoherence we can reevaluate for what ranges of parameters HM decay dominates over CDL decay, including environmental interactions. Using $\Gamma_{\te{CDL}}^{\te{dec}}\sim\Gamma_{\te{CDL}}^{2}$ and  $\Gamma_{\te{HM}}^{\te{dec}}\sim\Gamma_{\te{HM}}$ as argued above we find (in Planck units)
\be 
{t_{\te{HM}}^{\te{dec}}\over t_{\te{CDL}}^{\te{dec}}}=\exp\bls 8\pi^{2}\bl{16\over S_{1}^{2}}-{3\over V_{0}}-{3\over V_{1}} \br \brs,
\ee
which indicates that for $3/V_{0}+3/V_{1}>16/S_{1}^{2}$, HM tunneling is the dominant decay channel. The HM decay rate is not changed by decoherence, so any de Sitter vacuum will decay before its lifetime exceeds the limits set by Poincar{\'e} recurrence. 

\section{Conclusions}\label{conclusion}
We have demonstrated that the timescale of Coleman-de\,Luccia decay is highly dependent on external modes to which the tunneling scalar field is coupled. Choosing a generic model of a tunneling scalar field and photons coupled to gravity, we have shown that for the lowest metastable initial state even de Sitter radiation is sufficient to induce an efficient quantum Zeno effect that suppresses vacuum decay significantly. We exploited the fact that the environmental modes are not sourced by the tunneling field itself, so that we were able to model the bubble-photon interaction using an S-matrix approach. Not only did the use of external modes greatly simplify the problem, it was also a crucial ingredient for obtaining efficient decoherence. While Coleman-de\,Luccia decay is strongly suppressed, we found that Hawking-Moss decay is not as significantly affected by interactions with the environment. Thus, the lifetime of de Sitter space does not exceed the limits set by the Poincar{\'e} recurrence time, even when environmental interactions are included. In this work we considered the lowest metastable initial state of the effective wave function. In more general scenarios with other initial states and potentials the significance of decoherence may change dramatically and needs to be evaluated carefully.

The strong suppression of the vacuum decay rate has a broad range of possible implications. In this paper we discussed one specific model of coupling the tunneling field to environmental modes gravitationally. In more realistic cosmological models one expects a far richer pool of fields that couple more strongly to a nucleating new vacuum. We suggest that a far greater suppression of the vacuum decay rate is achievable in such scenarios, e.g.\ by considering couplings to dark matter or CMB photons. It would be interesting to characterize what the constraints on the stability of a de Sitter vacuum are when these decoherence effects are included. In particular, one might expect an effective decay rate that is increasing with time as the universe gets more and more dilute and decoherence loses efficiency.

\section{Acknowledgements}
I would like to thank Mathieu Cliche and Friedemann Queisser for useful discussions. I am particularly grateful to Liam McAllister for valuable discussions and comments on the draft. I would like to thank the organizers of the PITP summer school at the Institute for Advanced Study, where this project was initiated, for providing a stimulating environment. This work was supported in part by the NSF under grant PHY-0757868.

\bibliographystyle{unsrt}

\end{document}